\documentclass[10pt,preprintnumbers,nofootinbib]{revtex4-2}%
\usepackage[latin9]{inputenc}
\usepackage{amsmath}
\usepackage{amssymb}
\usepackage{graphicx}
\usepackage[colorlinks=true, pdfstartview=FitV, linkcolor=blue, citecolor=red, urlcolor=magenta]{hyperref}
\usepackage{amsfonts}
\usepackage{subfigure}
\usepackage{cleveref}
\usepackage{listings}
\usepackage{array}
\usepackage{url}
\usepackage{comment}
\usepackage{todonotes}

\usepackage{tensor}
\newcommand{\df}{\mathrm{d}}
\newcommand{\veps}{\bm{\epsilon}}

\newcommand{\LL}{\mathcal{L}}
\newcommand{\FF}{\mathcal{F}}
\newcommand{\GG}{\mathcal{G}}
\usepackage{xcolor}
\usepackage{soul}
\allowdisplaybreaks

\graphicspath{{Figures/}}

\begin{document}

\title{Noncommutative dyonic black holes sourced by nonlinear electromagnetic fields}
\author{Ana Bokuli\'c}
\email{ana.bokulic@ua.pt}
\affiliation{Departamento de Matem\'atica da Universidade de Aveiro and Centre for Research and Development  in Mathematics and Applications (CIDMA), Campus de Santiago, 3810-193 Aveiro, Portugal.}
\author{Filip Po\v{z}ar}
\email{filip.pozar@irb.hr}

\affiliation{Rudjer Bo\v{s}kovi\'c Institute, Bijeni\v cka  c.54, HR-10000 Zagreb, Croatia}

\begin{abstract}
We introduce the first-order noncommutative (NC) corrections to the general nonlinear electrodynamics (NLE) Lagrangian depending on two electromagnetic invariants. The NC deformation of Einstein-NLE theory is implemented using the $\partial_t\wedge\partial_\varphi$ Drinfel'd twist and the NC effects are encoded in the matter sector through the Seiberg-Witten map. The resulting equations of motion reflect two distinct sources of nonlinearity in this framework; one arising from replacing Maxwell's electrodynamics with its nonlinear modifications and another from the NC deformations. Assuming a general form of static, spherically symmetric dyonic black hole as a seed solution in the commutative limit, we solve the equations of motion perturbatively to the first order in the NC parameter $a$. Finally, we evaluate the obtained corrections to the metric tensor and gauge potential for several prominent NLE theories.
\end{abstract}
\maketitle
%\tableofcontents
\section{Introduction}
Black hole solutions in general relativity coupled to nonlinear electrodynamics (NLE) have long been a fertile ground for exploring modifications to classical singularity theorems and the structure of spacetime in the presence of strong electromagnetic fields. NLE theories extend the standard Maxwell electrodynamics by introducing a Lagrangian that is a general function of the electromagnetic invariants 
$\mathcal{F}=\tensor{F}{_a_b}\tensor{F}{^a^b}$ and $\mathcal{G}=\tensor{F}{_a_b}{\ast\tensor{F}{^a^b}}$, leading to nonlinear field equations. Early NLE theories were motivated by the attempts to regularise the divergent self-energy of point charges \cite{Born34, BI34} or emerged as effective theories within quantum field theory \cite{HE36}. In the subsequent years, the catalogue of NLE Lagrangians has expanded, driven by two main, often intertwined, objectives: constructing NLE theories with particular symmetries \cite{BLST20} and searching for novel spacetimes with specific physical or geometric properties. Despite challenges coming from the nonlinear nature of the matter sources, significant progress has been achieved within the family of static, spherically symmetric black holes. Among these, the most notable are regular black holes \cite{ABG00, BOKULIC2024138750}, whose construction often relies on introducing magnetic charge, a result substantiated by several no-go theorems \cite{Bronnikov00, BJS22}. Another important subclass consists of dyonic configurations, which provide a key testing ground for duality symmetries and the interplay between the two electromagnetic invariants, $\mathcal{F}$ and $\mathcal{G}$.\\

While NLE theories provide a way of exploring classical modifications to Einstein-Maxwell theory, noncommutative (NC) geometry, on the other hand, offers a framework to probe the conjectured quantum structure of spacetime at Planck scales. This is achieved by deforming the classical algebra of functions on a manifold into a noncommutative algebra. Among many approaches to this subject, one systematic approach exploits Drinfel'd twists of the vector field Hopf algebra and its modules, yielding star-product deformations of the differential geometry. This twisted formalism, introduced in \cite{Aschieri_2005, aschieri_kulish_wess_dimitrijevic_lizzi_2009} and outlined in \cite{Aschieri:2009qh,Juric:2022bnm,Herceg:2023pmc}, defines covariance under a deformed diffeomorphism symmetry and has been applied to scalar fields, gauge fields, and gravity in both pure and matter-coupled contexts. Some other approaches to noncommutative gauge theories include the exact Seiberg-Witten map \cite{Martin} (nonperturbative in the NC parameter), obtained by expanding only in the gauge coupling constant, and the braided gauge theories \cite{6df9564f199b43ceb14cac28eca9bf4c}.\\

For a long time, the only known solutions to Einstein's equations in this approach to the noncommutative geometry coincided with the commutative ones, having too much symmetry to produce any deviations from the commutative results \cite{Aschieri:2009qh, Ohl:2009pv}. In the absence of solutions to NC equations of motion, many works used various motivations to introduce NC corrections to known metrics. Some approaches include the NC Poincar\'{e} gauge fields \cite{Juric:2025kjl, AraujoFilho:2025jcu, Heidari:2025iiv}, the effective NC metric \cite{DimitrijevicCiric:2024ibc} or the NC inspired smearing of classical black holes \cite{Nicolini:2005vd}. Recently, though,
\begin{comment}
a symmetrized version of the noncommutative Ricci tensor was postulated as the NC Einstein equation and was applied to the perturbation problem of the Schwazschild black hole - producing different dynamics, parametrized by the NC parameter $a$, compared to the commutative Einstein equation. 
\end{comment}
in the work \cite{Pozar:2025yoj}, a very general family of noncommutative Einstein-Hilbert-Maxwell (EHM) actions was considered by promoting the differential geometry structure of EHM theory into the NC one. By employing the Palais' symmetric criticality theorem \cite{Palais:1979rca, Deser:2003up,Fels:2001rv,Torre:2010xa}, assuming axial and time symmetry of the fields, the NC EHM actions reduced to their shared commutative counterpart, but with the matter fields being expanded in the Seiberg-Witten map \cite{Seiberg:1999vs}. In this way, the initially complicated actions with many new terms proportional to the NC parameter $a$, coming from the Seiberg-Witten map and from the noncommutative product structure, get simplified. In fact, they are reduced to a shared tractable form differing from the commutative EHM action only by a few terms proportional to $a$. A time- and axially-symmetric solution was found, corresponding to the Kerr-Newman metric and its associated gauge potential, with nonzero corrections linear in $a$. These corrections can be completely traced back and attributed to the deformed/noncommutative gauge symmetry of the matter content of the EHM theory. An important point that was missed in \cite{Pozar:2025yoj} is that the coupled NC theory of gravity and electromagnetism can not achieve gauge invariance in general\footnote{Star products with the gravitational objects appear in between of the gauge transition functions $U$ and $U^{-1}$, obstructing their cancellation.}. Nevertheless, (NC) gauge transformations of the NC Kerr-Newman solution are still solutions to all of the starting NC actions. This suggests that the naive promotion of commutative products to star products and fields to SW expanded fields does not produce a very realistic action principle of the coupled theory, but, the static and axially symmetric solution might still encapsulate some true NC effects due to the emergence of the gauge symmetry of the solution.\\

Having recognized the important role of matter fields in generating nontrivial NC corrected metrics, in the manner of \cite{Pozar:2025yoj}, we will consider the effects of deforming the $U(1)$ gauge symmetry of nonlinear electromagnetic fields sourcing black hole spacetimes. More precisely, we will find dyonic black hole solutions in the framework of noncommutative geometry, such that the obtained NC generalizations will reproduce the commutative dyonic solutions in the commutative limit. By doing so, we will bridge the two areas of research: we will provide a novel type of nonlinearity (arising from the NC corrections to the NLE action) to the NLE community, while also offering new NC corrected solutions (sourced by NLE fields) to the noncommutative gravity community. Again, as in \cite{Pozar:2025yoj}, the deformation of the $U(1)$ symmetry is strong enough that the symmetry is lost in general, but persists when gauge transformations are applied to our solution with time and axial symmetry.\\

The paper is organized as follows : In sections \ref{sec2} and \ref{sec3}, we review NLE theories and NC geometry within the framework of Einstein's gravitational theory. Section \ref{sec3} also introduces the NC Einstein-NLE action, where the electromagnetic field tensor is expanded via the SW map. The NC-NLE equations of motion, valid for a Killing Drinfel'd twist $\partial_t\wedge\partial_\varphi$, are derived in section \ref{sec4}. In section \ref{sec5}, we obtain a general perturbative NC correction to the dyonic NLE black hole solution. We further comment on the implications of these corrections by examining specific NLE theories. Finally, section \ref{sec6} presents our conclusions and outlines possible directions for future research.

\section{Fundamentals of NLE}\label{sec2}
Before proceeding to analyze the combined effects of NC and NLE modifications, we present an overview of each area separately, starting from NLE theories.\\

To formulate an NLE theory, one typically begins by replacing Maxwell's Lagrangian density with a more general function of the electromagnetic invariants constructed from the electromagnetic 2-form $F_{ab}$ and its Hodge dual, $\ast\tensor{F}{_a_b}=\tensor{\epsilon}{_a_b_c_d}\tensor{F}{^c^d}/2$. In this work, we will focus on Lagrangians that depend on two quadratic electromagnetic invariants, denoted by
\begin{align}
\mathcal{F}=\tensor{F}{_a_b}\tensor{F}{^a^b}\ \  \text{and}\ \ \mathcal{G}=\tensor{F}{_a_b}{\ast\tensor{F}{^a^b}}\ .
\end{align}
In fact, it can be shown \cite{EU14} that scalars formed by contracting an arbitrary number of 2-forms $\tensor{F}{_a_b}$ and ${\ast \tensor{F}{_a_b}}$ reduce to combinations of higher powers of these quadratic invariants. While a broader approach could consider invariants that include covariant derivatives of $F_{ab}$, our restriction still encompasses the majority of models explored in the literature. Furthermore, we will consider minimal coupling between gravitational and electromagnetic fields.
With these assumptions, the total Lagrangian 4-form takes the form
\begin{align}\label{eq:NLEact}
\textbf{L}=\dfrac{1}{16\pi}(R+4\mathcal{L}(\mathcal{F},\mathcal{G}))\veps\ ,
\end{align}
where the gravitational part consists of the Einstein-Hilbert term and $\mathcal{L}(\mathcal{F},\mathcal{G})$ is the NLE contribution. The (commutative) Einstein-NLE equation of motion derived from (\ref{eq:NLEact}) via the variational principle is
\begin{align}\label{eq:nlemt}
\tensor{R}{_a_b}-\dfrac{1}{2}R\tensor{g}{_a_b}=8\pi\tensor{T}{_a_b}\ ,
\end{align}
with the energy momentum tensor given by
\begin{align}
\tensor{T}{_a_b}=-\dfrac{1}{4\pi}((\mathcal{L}_\mathcal{G}\mathcal{G}-\mathcal{L})\tensor{g}{_a_b}+4\mathcal{L}_\mathcal{F}\tensor{F}{_a_c}\tensor{F}{_b^c})\ .
\end{align}
Here we have introduced the shorthand notation for derivatives of the Lagrangian density, $\mathcal{L}_\mathcal{F}=\partial_\mathcal{F}\mathcal{L}$ and $\mathcal{L}_\mathcal{G}=\partial_\mathcal{G}\mathcal{L}$.
After defining the auxiliary 2-form 
\begin{align}
\tensor{Z}{_a_b}=-4(\mathcal{L}_\mathcal{F}\tensor{F}{_a_b}+\mathcal{L}_\mathcal{G}{\ast\tensor{F}{_a_b}})\, ,
\end{align}
the generalized Maxwell's equations can be written compactly as
\begin{align}\label{eq:NLEMax}
\df{\textbf{F}}=0 \ \ \text{and}\ \ \df{\ast \textbf{Z}}=0\ .
\end{align}
The expressions for the charges in the NLE theories have to be modified accordingly: Komar integrals for the electric charge $Q$ and magnetic charge $P$ are defined as
\begin{align}\label{eq:Komar}
Q=\dfrac{1}{4\pi}\oint_\mathcal{S} {\ast\textbf{Z}}\ \  \text{and}\ \  P=\dfrac{1}{4\pi}\oint_\mathcal{S}\textbf{F}\ ,
\end{align}
where $\mathcal{S}$ denotes a closed 2-surface. The 2-form $\tensor{F}{_a_b}$ can be decomposed with respect to a non-null vector field $X^a$ as 
\begin{align}-(X^cX_c)\textbf{F}=\textbf{X}\wedge\textbf{E}+\ast(\textbf{X}\wedge\textbf{B})\ ,
\end{align}
where electric and magnetic forms are given by
\begin{align}\label{eq:fields}
\textbf{E}=-i_X\textbf{F}\ \ \text{and}\ \ \textbf{B}=i_X{\ast{\textbf{F}}}\ .
\end{align}
One may also introduce two additional forms using the auxiliary form $\tensor{Z}{_a_b}$,
\begin{align}
\textbf{D}=-i_X\textbf{Z}\ \ \text{and}\ \ \textbf{H}=i_X{\ast{\textbf{Z}}}\ .
\end{align}
The most convenient choice for static spacetimes is the decomposition with respect to the Killing vetor field $\partial_t$. It can be shown that, assuming symmetry inheritance of the electromagnetic field \footnote{The analysis of symmetry inheritance in the case of NLE fields was presented in \cite{PhysRevD.95.124037}}, forms $\textbf{E}$ and $\textbf{H}$ are closed and therefore admit globally defined scalar potentials on simply connected domains \cite{BJS21}.\\
Apart from satisfying the superposition principle, Maxwell's electrodynamics exhibits two important properties: duality invariance and conformal symmetry. These symmetries also simplify the task of solving the coupled Einstein-Maxwell equations. Therefore, NLE theories that share either of the aforementioned symmetries are of particular interest. Conformal symmetry is characterised by a vanishing trace of the energy momentum tensor, i.e., $\mathcal{L}-\mathcal{L}_\mathcal{F}\mathcal{F}-\mathcal{L}_\mathcal{G}\mathcal{G}=0$,  while duality invariance requires $\mathcal{G}-{\ast{\tensor{Z}{_a_b}}}\tensor{Z}{^a^b}=0$ \cite{GR95}. From a physical standpoint, NLE theories that reduce to Maxwell's electrodynamics in the weak field limit are especially relevant. To make the statement precise, Maxwell's weak field (MWF) limit implies $\mathcal{L}_\mathcal{F}\to-1/4$ and $\mathcal{L}_\mathcal{G}\to0$ as $(\mathcal{F},\mathcal{G})\to(0,0)$. For a more comprehensive review of NLE theories, see the references \cite{osti_4071071, https://doi.org/10.1002/prop.202200092}.

\section{Noncommutative Einstein-NLE action}\label{sec3}
In this section we will outline some important features of NC geometry and derive the NC Einstein-NLE action for the $\partial_t\wedge\partial_\varphi$ twist in a special symmetrical regime by emplyoing the Palais' theorem. Let us proceed by first considering NC deformations of a large family of Einstein-NLE actions \eqref{eq:NLEact}. By a theory's NC deformation we mean writing its Lagrangian in terms of the NC differential geometric structure, i.e., replacing commutative $\cdot$ products in the component contractions by the $\star$ products and replacing commutative gauge fields with NC gauge fields obeying NC gauge transformations. As mentioned in the Introduction, the approach to noncommutative geometry that we consider in this paper is the twisted Hopf algebra approach, which was outlined in \cite{Aschieri:2009qh,Juric:2022bnm,Herceg:2023pmc} and applied in \cite{Pozar:2025yoj}, which resulted in the first nontrivial\footnote{In the sense that the NC solution was not just equal to the commutative solution, as was the case in \cite{Aschieri:2009qh, Ohl:2009pv}, but it also had nonzero correction terms} NC gravitational black hole solution. In summary, the canonical Hopf algebra of vector fields $\left(\mathcal{U}(\Xi), m, \Delta, \epsilon, S\right)$ along with its modules completely describes the differential geometry of a smooth manifold. Then, using a Drinfeld element $\mathcal{F}\in \mathcal{U}(\Xi)\otimes\mathcal{U}(\Xi)$, it is possible to systematically deform this Hopf algebra's coproduct and antipod, turning it into a noncocommutative Hopf algebra. As a byproduct, the modules on which this deformed Hopf algebra acts are given as deformations of the commutative Hopf algebra's modules. The multiplication structure in the deformed Hopf algebra's modules can be expressed as
\begin{equation}
    a_1\star a_2 = \cdot \circ \mathcal{F}^{-1}\left(a_1\otimes a_2\right)
\end{equation}
where $\star$ is the deformed multiplication and $\cdot$ is the multiplication in the undeformed module (of the underformed Hopf algebra). This immediately implies that the module of complex valued manifold functions becomes noncommutative. For the Moyal twist
\begin{equation}
    \mathcal{F} = e^{-i\frac{a}{2}\theta^{\alpha\beta}\partial_\alpha\otimes \partial_\beta}\, ,\quad\theta^{\alpha\beta} = - \theta^{\beta\alpha}\;,
\label{Moyal}
\end{equation}
which we will use exclusively in our paper, we obtain the well-known Moyal product of functions
\begin{equation}
    \left(f\star g\right)(x) = \lim_{x\rightarrow y}e^{i\frac{a}{2}\theta^{\alpha\beta}\partial_\alpha \partial_\beta} f(x) g(y) = f(x)g(x) + i\frac{a}{2}\theta^{\alpha\beta}\partial_\alpha f(x) \partial_\beta g(x) + \mathcal{O}(a^2).
\end{equation}
For the antisymmetric matrix $\theta$ we consider the one generating the so called $\partial_t\wedge\partial_\varphi$ twist,
\begin{equation}
\theta^{{\alpha\beta}} \partial_\alpha \otimes \partial_\beta = \partial_t\wedge\partial_\varphi\;,
\label{tphi twist}
\end{equation}
which is especially nicely suited for axially symmetric static spacetimes. The parameter $a$ in the Moyal twist element \eqref{Moyal} is called the noncommutative parameter and it is treated as being of the scale of Planck length $l_{\text{Pl}}$. That is because in the limit $a\rightarrow 0$, called the commutative limit, the deformed Hopf algebra and all of its modules revert back to the usual differential geometry scenario. Since the noncommutativity of spacetime, if at all existing, is thought to be relevant at the Planck length scales, it is natural to impose this scale to the parameter $a$.\\

Having said all of this, we can now write examples of some NC action principles in the language of Moyal twists using the Moyal product, e.g.,
\begin{equation}
    S[\phi] = \int d^4x\sqrt{-g}\star\left[  \phi^\dagger\star \phi - \partial_\mu \phi \star \partial^\mu \phi^\dagger \right]
\label{Sphi}
\end{equation}
is an action functional which would be natural to consider as a free NC scalar field theory in curved background. A complication that should not be dismissed is that also, e.g., the action 
\begin{equation}
    S'[\phi] = \int d^4x\left[  \phi^\dagger\star\sqrt{-g}\star \phi -  \partial^\mu \phi^\dagger\star\partial_\mu\phi\star\sqrt{-g} \right]
\end{equation}
is equally motivated to be considered a free NC scalar field theory in curved background, but has different equations of motion compared to \eqref{Sphi}. In a similar manner, by taking convex combinations of such actions, one could find infinitely many equally motivated NC generalizations of a given action principle in commutative geometry, which all reproduce the commutative action
\begin{equation}
    S_c[\phi] = \int d^4x \sqrt{-g}\left[\phi^\dagger \cdot\phi - \partial_\mu \phi \cdot\partial^\mu\phi^\dagger\right]
\end{equation}
in the commutative scenario $\star\mapsto \cdot$. The ordering issue is crucial for theories that aim to make phenomenological predictions, as the solutions to the equations of motion depend on the specific ordering of $\star$ products at every order $a^n$ for $n\geq 1$. In this section and the following one, we will show that it is actually possible to consider all such theories simultaneously and to find a solution which they all share, avoiding the ambiguities associated with the ordering and vastly increasing the scope of studying solutions to NC deformed actions.\\

But, before proceeding, it is important to notice that generalizing gauge theories is not so straightforward. Namely, the commutative gauge transformations
\begin{equation}
    A_\mu \mapsto A_\mu + \delta_\lambda A_\mu = A_\mu + \partial_\mu \lambda + \left[A_\mu,\lambda\right]
\label{comm gauge}
\end{equation}
are not defined in NC geometry. Exponentiating an infinitesimal transformation is not possible because the standard exponential, given by a power series of commutative compositions, is ill-defined in this context. Similarly, also the commutator term in \eqref{comm gauge} is not defined in NC geometry, which only has $\star$ products. Having noticed this, it is neccesary to generalize connections and their curvatures to the noncommutative geometry scenario. This was heavily studied in mathematical physics literature, for one of many reviews see \cite{Hersent:2022gry}. What turns out to be a good notion of a noncommutative gauge field (denoted with a hat $\hat{}$\;) is a field obeying the NC gauge transformation rule
\begin{equation}
    \hat{A}_\mu \mapsto U\star \hat{A}_\mu\star U^{-1} +
     U^{-1}\star \partial_\mu U\;,
\end{equation}
for
\begin{equation}
    U^{\pm 1} = \sum _{n=0}^\infty \frac{1}{n!}(\pm i \hat{\lambda})^{\star n}\;,
\end{equation}
with $\hat{\lambda}$ a NC gauge parameter. Additionally, the components of the NC curvature $\hat{F}_{\mu\nu}$ on Moyal spacetime turn out to be
\begin{equation}
    \hat{F}_{\mu\nu} = \partial_\mu \hat{A}_\nu - \partial_\nu \hat{A}_\mu - i\left[\hat{A}_\mu,\hat{A}_\nu\right]_\star \equiv \partial_\mu \hat{A}_\nu - \partial_\nu \hat{A}_\mu - i\left( \hat{A}_\mu\star\hat{A}_\nu - \hat{A}_\nu \star \hat{A}_\mu\right)
\end{equation}
which transform as
\begin{equation}
    \hat{F}_{\mu\nu}\mapsto U\star \hat{F}_{\mu\nu}\star U^{-1}
\end{equation}
under gauge transformations. However, since the NC exponential, given as a power series with powers defined using $\star$ products, can be expressed as the commutative exponential plus $a$ proportional corrections, and since the NC $\star$ commutator is also given as the usual commutator plus $a$ proportional corrections, sometimes it is actually possible to define, order by order in $a$, a link between commutative and noncommutative gauge objects (connection, curvature and gauge parameter). The obstructions to this link were discussed in \cite{Kraus:2001xt, Salizzoni:2005eh} where it was found that, at finite $a$, topological solutions can exist which do not have a well defined commutative limit. In this paper we will only consider noncommutative gauge potentials which have well defined commutative limits\footnote{We are interested in corrections due to spacetime noncommutativity in regimes where noncommutative geometry is a small effective correction, not dominating behavior.} $a\rightarrow 0$ for which the mentioned link exists and is realized as the famous Seiberg-Witten map \cite{Seiberg:1999vs}. The Seiberg-Witten (SW) map of the gauge field $A_\mu$ produces the NC gauge field $\hat{A}_\mu$ given as
\begin{equation}\label{ASW}
    \hat{A}_\mu = A_\mu -\frac{aq}{2}A_\alpha\left(\partial_\beta A_\mu + F_{\beta\mu}\right)\theta^{\alpha\beta} + \mathcal{O}(a^2)\;.
\end{equation}
The SW map also defines a mapping between commutative gauge parameters $\lambda$ and NC gauge parameters $\hat{\lambda}$,
\begin{equation}
    \hat{\lambda} = \lambda + \frac{aq}{2}\theta^{\alpha\beta} \partial_\alpha\lambda A_\beta + \mathcal{O}(a^2)\, ,
\end{equation}
in such a way that the gauge transformed $A_\mu$ maps to the NC gauge transformed $\hat{A}_\mu$
\begin{equation}
    A_\mu + \delta_\lambda A_\mu \mapsto \hat{A}_\mu + \hat{\delta}_{\hat{\lambda}}\hat{A}_\mu\;.
\end{equation}
The NC curvature with a well defined commutative limit can also be obtained by the following SW map
\begin{equation}
    \hat{F}_{\mu\nu} = F_{\mu\nu} -\frac{aq}{2} \theta^{\alpha\beta}A_\alpha\left(\partial_\beta F_{\mu\nu} + D_\beta F_{\mu\nu}\right) + aq\; \theta^{\rho\sigma}\tensor{F}{_\rho_\mu}\tensor{F}{_\sigma_\nu} + \mathcal{O}(a^2)\;,
\label{SW F}
\end{equation}
where $D_\beta F_{\mu\nu}=\partial_\beta F_{\mu\nu}$ for U(1) electromagnetic theory. One may notice the coupling constant $q$ appearing in all Seiberg-Witten maps. This coupling parameter is intrinsic to the Seiberg-Witten map as can be seen in the original paper \cite{Seiberg:1999vs}, where it is absorbed\footnote{In the original paper \cite{Seiberg:1999vs}, the SW map is expressed in terms of the gauge field $A^{(\mathrm{SW})} \equiv qA$ with the parameter $q$ absorbed.} into the gauge field $A$. The parameter $q$ is dimensionful and for theories with matter content, it is often identified as the charge of the matter field \cite{Juric:2022bnm, Ciric:2017rnf, Herceg:2025zkk}. In our considerations, we can not fix the value of $q$ by any criterion, but, the parameter $q$ will always multiply the NC parameter $a$ so we can simply regard $aq$ as the effective NC parameter. Finally, although not used in this paper, the SW map also exists for fields transforming in the vector representation of the gauge group. In the rest of this paper, we will absorb the effective noncommutative parameter $aq$ into the Moyal matrix $\theta^{\mu\nu}$, writing it explicitly again only in the final results. \\

Now we are ready to proceed with the consideration of NC Einstein-NLE theories. Take any action of the form
\begin{equation}
    S = \int d^4x \sqrt{-g} \Bigl[\mathcal{L}^*_{GRAV}(g) + \mathcal{L}^*_{NLE}(g,\hat{A})\Bigr]
\label{nc action}
\end{equation}
obtained by writing the action \eqref{eq:NLEact} with $\star$ products, choosing the orderings of contractions and by promoting the gauge field $A_\mu$ into $\hat{A}_\mu$ (and its curvature $F_{\mu\nu}$ into $\hat{F}_{\mu\nu}$). As mentioned before, the NC $\star$ product and SW map are obtained using the Moyal twist generated by the antisymmetric matrix $\theta^{\alpha\beta}$, whose only nonzero entries are for $(\alpha,\beta) = (t,\varphi)$ and for $(\alpha,\beta) = (\varphi,t)$, as defined in \eqref{tphi twist}. In a general case, one can not make any statements without choosing $\mathcal{L}^*_{GRAV}$ and $\mathcal{L}^*_{NLE}$, but for static and axially symmetric solutions we can employ Palais' principle of symmetric criticality \cite{Palais:1979rca}, in a similar way as was done in \cite{Deser:2003up}, to impose the $\partial_{\varphi}$ and $\partial_t$ symmetry of fields in the action\footnote{The Palais' principle is a theorem which states that when looking for solutions with some spacetime symmetry, it is allowed to impose the symmetry on the level of action, vary it and look for such symmetric solutions. The results will match to the symmetric solutions of the starting unmodified action.}. For the Killing twist \eqref{tphi twist}, if at least one field in the $\star$ product does not depend on $t$ and $\varphi$, the $\star$ product equals the commutative pointwise product between functions. This means that imposing the axial and time symmetry on both the metric and the electromagnetic potential will very strongly impact the action, putting it into the form
\begin{equation}
S = \frac{1}{16\pi}\int d^4x \sqrt{-g}\Bigl[R + 4\mathcal{L}(\hat{\mathcal{F}}, \hat{\mathcal{G}})\Bigr]\;.
\label{NC NLE}
\end{equation}
Since the Palais theorem requires the symmetry restriction to be applied to all fields appearing in the action, this step was not merely consistent with the theorem but in fact required by it. In other words, the static and axially symmetric solutions of \eqref{NC NLE} are the same as static, axially symmetric solutions of any action of the form \eqref{nc action} in the $\partial_t\wedge\partial_{\varphi}$ twist. A natural question which arises is why does \eqref{NC NLE} need Seiberg-Witten expanded gauge fields, instead of simply commutative gauge fields, if the action \eqref{NC NLE} does not contain $\star$ products? The reason is that \eqref{NC NLE} is simply a restriction of \eqref{nc action} to axially symmetric and static fields, but \eqref{NC NLE} needs to transform under NC gauge transformations for any NC gauge parameter $\hat{\lambda}(x)$, including those that depend on $t$ and $\varphi$ nontrivially. To account for such NC gauge parameters, the SW map needs to remain in \eqref{NC NLE}. All in all, we refer to the action \eqref{NC NLE} as the NC Einstein-NLE action, which applies to solutions where the Drinfeld twist is Killing. One also arrives at the same NC Einstein-NLE action \eqref{NC NLE} in the semi-Killing scenario, where only one coordinate in the twist is Killing for all the fields.
\\
\subsection{Gauge invariance of Killing solutions}\label{Subsec 3a}
Finally, as mentioned in the Introduction, the action \eqref{nc action} is not gauge invariant in general. In this Subsection we will show how none of the NLE theories coupled to gravity, in the sense in which we introduced them, are gauge invariant. But, we will also show that NC gauge transformations of our Killing solutions are still solutions to any action \eqref{nc action}. For example, in the case of Maxwell's electrodynamics coupled to gravity\footnote{Here we omit discussing the difficulties of what the determinant or the square root in NC geometry is, which is relevant for the $\sqrt{-g}$ term. Namely, regardless of the definition, the same problem regarding the gauge invariance will appear.} with the orderings (in the electromagnetic sector) as follows
\begin{equation}
    S \supset \int d^4x \sqrt{-g}\star \hat{F}_{\mu\nu}\star \hat{F}_{\rho\sigma} \star g^{\mu\rho}\star g^{\nu\sigma}\, ,
\label{maxwell + gravity}
\end{equation}
the gauge transformation is given as
\begin{equation}
    S^g \supset \int d^4x \sqrt{-g}\star U \star \hat{F}_{\mu\nu}\star U^{-1}\star U\star  \hat{F}_{\rho\sigma} \star U^{-1}\star g^{\mu\rho}\star g^{\nu\sigma} = \int d^4x \sqrt{-g}\star U \star \hat{F}_{\mu\nu}\star  \hat{F}_{\rho\sigma} \star U^{-1}\star g^{\mu\rho}\star g^{\nu\sigma}\;.
\end{equation}
Now, even the trace cyclicity of the integral with respect to the star product
\begin{equation}
\label{trace cyclicity}
    \int d^4 x\; f\star g \star h = \int d^4x\; h\star f \star g \quad,\ \  \forall f,g,h \in C^\infty(M)
\end{equation}
is not enough to assure gauge invariance, because the $U$ and $U^{-1}$ in general do not commute with the metrics and as such can not cancel. On the other hand, notice that if the metric $g$ lies in the part of the configuration space which is invariant (Killing) with respect to both vector fields appearing in the Moyal twist, the metric $\star$-commutes with every function in $C^\infty(M)$, with the product reducing simply to the pointwise product. This means that the gauge transformation of the action \eqref{maxwell + gravity} becomes
\begin{equation}
    S^g \supset\int d^4x \sqrt{-g} g^{\mu\rho} g^{\nu\sigma} U \star \hat{F}_{\mu\nu}\star \hat{F}_{\rho \sigma} \star U^{-1}\;,
\end{equation}
which, using the trace cyclicity \eqref{trace cyclicity}, gives back \eqref{maxwell + gravity}. Identical arguments hold for any NLE Lagrangian given as an analytical function of $\mathcal{F}$ and $\mathcal{G}$ around $\mathcal{F}=0,\ \mathcal{G}=0$,
\begin{equation}
    \mathcal{L} = \sum_{n,m = 0}^\infty c_{n,m} \mathcal{F}^n\mathcal{G}^m\;.
\label{LNLE taylor}
\end{equation}
Having established this, the solution of the equations of motion derived from the action \eqref{nc action} supplied by Palais' theorem contains a metric that is Killing with respect to both vector fields appearing in the Moyal twist. We have shown that when the action is evaluated on such a Killing metric, it remains gauge invariant, independently of whether the gauge potential itself satisfies any Killing condition.
Additionally, gauge transformations are simply translations by $\partial_\mu \alpha$ in the $A_\mu$ configuration space, so entire open neighbourhoods of the Killing configuration get translated into open neighbourhoods of the translated Killing configuration. It follows that the gauge transformation of the Killing configuration (which need not be Killing any more after gauge transformation) retains the status of a local extremum.
The final step of this proof relies on restricting the configuration space to NC gauge fields which can be expressed as SW maps of commutative fields, i.e., they have well-defined commutative limits. Then, we can search for commutative solutions of the SW-expanded action, without worrying that an NC gauge transformation might map the Killing configuration into a neighbourhood containing gauge field configurations that are nonperturbative (akin to the ones in \cite{Kraus:2001xt, Salizzoni:2005eh}) in the deformation parameter $a$, and which could potentially be more extremal than the transformed Killing solution.
\section{NC Einstein-NLE equations}\label{sec4}
For the sake of generality, we will write the full $\mathcal{O}(a^1)$ expansion of the action \eqref{NC NLE}, without specifying the twist. In order to do so, first we need expansions of the electromagnetic invariants,
\begin{equation}
\begin{split}
    \hat{\mathcal{F}} &= \hat{F}_{\mu\nu}\hat{F}^{\mu\nu} = \mathcal{F} + g^{\mu\rho}g^{\nu\sigma}\hat{F}_{\rho\sigma}^{(1)}F_{\mu\nu} + g^{\mu\rho}g^{\nu\sigma}F_{\rho\sigma}\hat{F}_{\mu\nu}^{(1)} \\
    &= \mathcal{F} + 2\theta^{\alpha\beta}g^{\mu\rho}g^{\nu\sigma}F_{\rho\sigma}F_{\alpha\mu}F_{\beta\nu} - 2\theta^{\alpha\beta}g^{\mu\rho}g^{\nu\sigma}A_\alpha \left(\partial_\beta F_{\mu\nu} \right)F_{\rho\sigma}\;,
\end{split}
\label{Fcorr}
\end{equation}
and 
\begin{equation}
\begin{split}
\hat{\mathcal{G}} &= \hat{F}_{\mu\nu}\left(*\hat{F}^{\mu\nu}\right) = \mathcal{G} + \frac{1}{2}\epsilon^{\mu\nu\rho\sigma}\hat{F}_{\rho\sigma}^{(1)} F_{\mu\nu} + \frac{1}{2}\epsilon^{\mu\nu\rho\sigma}F_{\rho\sigma}\hat{F}_{\mu\nu}^{(1)} \\
&= \mathcal{G} + \theta^{\alpha\beta}\epsilon^{\mu\nu\rho\sigma}F_{\alpha\mu}F_{\beta\nu}F_{\rho\sigma} - \theta^{\alpha\beta}\epsilon^{\mu\nu\rho\sigma}A_{\alpha}\left(\partial_\beta F_{\mu\nu}\right) F_{\rho\sigma}\; ,
\end{split}
\label{Gcorr}
\end{equation}
where $\hat{F}_{\mu\nu}^{(1)}$ denotes terms in the SW expansion \eqref{SW F} proportional to the NC parameter $a$. Here the $A_{\mu}$ and $F_{\mu\nu}$, as before, represent the commutative limit of the gauge potential and its curvature. Focusing on the first-order correction in $a$ of the action \eqref{NC NLE}, using \eqref{Fcorr} and \eqref{Gcorr}, we get 
\begin{equation}
\begin{split}
S=\dfrac{1}{16\pi}\int\Biggl[R + 4\Bigl(&\mathcal{L}(\mathcal{F},\mathcal{G}) + 2\mathcal{L}_{\mathcal{F}}\theta^{\alpha\beta}g^{\mu\rho}g^{\nu\sigma}\left(F_{\alpha\mu}F_{\beta\nu} - A_\alpha \partial_\beta F_{\mu\nu} \right)F_{\rho\sigma}+ \\
&\mathcal{L}_{\mathcal{G}}\theta^{\alpha\beta}\epsilon^{\mu\nu\rho\sigma}\left(F_{\alpha\mu}F_{\beta\nu}- A_\alpha \partial_\beta F_{\mu\nu} \right)F_{\rho\sigma}\Bigr)\Biggr]\sqrt{-g}\,\df^4x\;.
\end{split}
\label{expanded action}
\end{equation}
The action \eqref{expanded action} has two sources of nonlinearity: the commutative NLE Lagrangian itself and the $\theta$ proportional corrections arising from the NC deformation of the gauge symmetry. In the Killing case, with the $\partial_t\wedge\partial_\varphi$ twist and time- and axially-symmetric fields, the terms proportional to $A_\alpha\partial_\beta F_{\mu\nu}$ will not contribute to equations of motion because we have assumed
\begin{equation}
    \partial_t \Psi = \partial_\varphi \Psi = 0
\end{equation}
for all fields $\Psi \in \left\{g_{\mu\nu}, A_\mu\right\}$ in the action. Had we chosen, e.g., the $\partial_\theta\wedge\partial_\varphi$ or $\partial_r\wedge\partial_\varphi$ twist and only axial symmetry of the fields (which would correspond to the semi-Killing situation), the discussed term would, in fact, contribute to equations of motion for $g$ and $A$. \\

Specializing to the Killing scenario $\partial_{t}\wedge\partial_{\varphi}$, we can now vary the action \eqref{expanded action} with respect to $g_{\mu\nu}$ and $A_\mu$ to obtain NC Einstein-NLE equations, keeping in mind that any solution that we find has to be invariant with respect to both Killing vectors $\partial_t$ and $\partial_\varphi$. Such solutions are the common solutions for all actions \eqref{nc action} in the $\partial_t\wedge\partial_\varphi$ twist. Up to $\mathcal{O}(a^1)$ order, for the Einstein equation we obtain
\begin{equation}
    G_{\mu\nu} = 8\pi \hat{T}_{\mu\nu}\;,
\label{NC efe}
\end{equation}
where $\hat{T}$ is the NC energy-momentum tensor given as
\begin{equation}\label{eq:NCEM}
\begin{split}
\tensor{\hat{T}}{_\mu_\nu}&=\dfrac{1}{4\pi}\Big(\tensor{g}{_\mu_\nu}\LL-4\LL_\mathcal{F}\tensor{F}{_\mu_\sigma}\tensor{F}{_\nu^\sigma}-\LL_\mathcal{G}\mathcal{G}\tensor{g}{_\mu_\nu}
+2\LL_\mathcal{F}\tensor{\theta}{^\alpha^\beta}\Big(\frac{1}{2}\tensor{F}{_\alpha_\beta}\FF+\frac{1}{4}\GG{\ast\tensor{F}{_\alpha_\beta}}\Big)\tensor{g}{_\mu_\nu}-4\LL_\mathcal{F}\tensor{\theta}{^\alpha^\beta}(\tensor{F}{_\alpha_\mu}\tensor{F}{_\nu_\sigma}+\tensor{F}{_\mu_\sigma}\tensor{F}{_\alpha_\nu})\tensor{F}{_\beta^\sigma}\\&-8\LL_{\mathcal{F}\mathcal{F}}\tensor{\theta}{^\alpha^\beta}\tensor{F}{_\mu_\tau}\tensor{F}{_\nu^\tau}\Big(\frac{1}{2}\tensor{F}{_\alpha_\beta}\FF+\frac{1}{4}\GG{\ast\tensor{F}{_\alpha_\beta}}\Big)-2\LL_{\mathcal{F}\mathcal{G}}\tensor{\theta}{^\alpha^\beta}\mathcal{G}\tensor{g}{_\mu_\nu}\Big(\frac{1}{2}\tensor{F}{_\alpha_\beta}\FF+\frac{1}{4}\GG{\ast\tensor{F}{_\alpha_\beta}}\Big)-2\LL_{\mathcal{F}\mathcal{G}}\tensor{\theta}{^\alpha^\beta}\tensor{F}{_\alpha_\beta}\GG\tensor{F}{_\mu_\tau}\tensor{F}{_\nu^\tau}\\&-\frac{1}{2}\LL_{\mathcal{G}\mathcal{G}}\tensor{\theta}{^\alpha^\beta}\tensor{F}{_\alpha_\beta}\GG^2\tensor{g}{_\mu_\nu}\Big)\, .
\end{split}  
\end{equation}
It can be easily seen that it is given as the commutative NLE energy-momentum tensor corrected by some $a^1$ terms. The calculation of the variation was carried out using the following auxiliary results;
\begin{align}
\delta\mathcal{F}=2\tensor{F}{_a_c}\tensor{F}{_b^c}\delta \tensor{g}{^a^b},\ \ \delta\mathcal{G}&=2\tensor{F}{_a_c}{\ast\tensor{F}{_b^c}}\delta\tensor{g}{^a^b},\ \ \delta\sqrt{-g}=-\dfrac{1}{2}\sqrt{-g}\tensor{g}{_a_b}\delta \tensor{g}{^a^b}\ .
\end{align}
To present the energy-momentum tensor \eqref{eq:NCEM} in a more tractable form, which will prove to be illuminating for the Killing twist scenario, we made use of the two identities valid for any 2-form,
\begin{align}
\tensor{F}{_a_c}{\ast\tensor{F}{^c_b}}={\ast\tensor{F}{_a_c}}\tensor{F}{^c_b}&=-\dfrac{1}{4}\mathcal{G}\tensor{g}{_a_b}\, ,\\
\tensor{F}{_a_c}\tensor{F}{^c_b}-{\ast\tensor{F}{_a_c}}{\ast\tensor{F}{^c_b}}&=-\dfrac{1}{2}\mathcal{F}\tensor{g}{_a_b}\, .
\end{align}
For the NC Maxwell equation, up to $a^1$ we have
\begin{equation}
\begin{split}\label{NC max}
\partial_\mu(\sqrt{-g}(&4\LL_\mathcal{F}\tensor{F}{^\mu^\nu}+4\LL_\mathcal{G}{\ast\tensor{F}{^\mu^\nu}}+4\LL_{\mathcal{F}}\tensor{\theta}{^\alpha^\beta}\tensor{g}{^\rho^\mu}\tensor{g}{^\sigma^\nu}\tensor{F}{_\alpha_\rho}\tensor{F}{_\beta_\sigma}+4\LL_{\mathcal{F}}\tensor{\theta}{^\mu^\beta}\tensor{F}{^\nu^\sigma}\tensor{F}{_\beta_\sigma}-4\LL_{\mathcal{F}}\tensor{\theta}{^\nu^\beta}\tensor{F}{^\mu^\sigma}\tensor{F}{_\beta_\sigma}\\+&4\LL_{\mathcal{G}}\tensor{\theta}{^\mu^\beta}{\ast\tensor{F}{^\nu^\sigma}}\tensor{F}{_\beta_\sigma}-4\LL_{\mathcal{G}}\tensor{\theta}{^\nu^\beta}{\ast\tensor{F}{^\mu^\sigma}}\tensor{F}{_\beta_\sigma}+2\LL_{\mathcal{G}}\tensor{\theta}{^\alpha^\beta}\tensor{\epsilon}{^\rho^\sigma^\mu^\nu}\tensor{F}{_\alpha_\rho}\tensor{F}{_\beta_\sigma}+8\LL_{\mathcal{F}\mathcal{F}}\tensor{\theta}{^\alpha^\beta}\tensor{F}{^\mu^\nu}\Big(\frac{1}{2}\tensor{F}{_\alpha_\beta}\FF+\frac{1}{4}\GG{\ast\tensor{F}{_\alpha_\beta}}\Big)+\\+&8\LL_{\mathcal{F}\mathcal{G}}\tensor{\theta}{^\alpha^\beta}{\ast\tensor{F}{^\mu^\nu}}\Big(\frac{1}{2}\tensor{F}{_\alpha_\beta}\FF+\frac{1}{4}\GG{\ast\tensor{F}{_\alpha_\beta}}\Big)+2\LL_{\mathcal{F}\mathcal{G}}\GG\tensor{\theta}{^\alpha^\beta}\tensor{F}{_\alpha_\beta}\tensor{F}{^\mu^\nu}+2\LL_{\mathcal{G}\mathcal{G}}\GG\tensor{\theta}{^\alpha^\beta}{\ast\tensor{F}{^\mu^\nu}}\tensor{F}{_\alpha_\beta})=0
\end{split}
\end{equation}
When deriving the NC Maxwell's equation, we are dealing with non-tensorial objects (with respect to diffeomorphism transformations) due to the presence of the $\theta^{\alpha\beta}$ matrix. For this reason, we used the partial derivative instead of the covariant, which has a well-defined defined action only on tensors.
It is important to notice that, since the equations of motion themselves are perturbative expansions in $a$, they must be solved order by order around the commutative solutions. This implies that the commutative terms appearing in \eqref{NC efe}, \eqref{NC max} also need to be expanded in $a$, since the commutative quantities in the EOM are evaluated on NC corrected metrics and potentials. For example, in the energy-momentum tensor $\hat{T}_{\mu\nu}$, the commutative contribution $-\frac{1}{2}g_{\mu\nu}\mathcal{L}$ must be evaluated on the corrected fields,
\begin{equation}
\begin{split}
g_{\mu\nu} &= g^{(0)}_{\mu\nu} +a h_{\mu\nu}\, ,\\
A_{\mu} &= A^{(0)}_\mu + a B_\mu\, ,
\end{split}
\end{equation}
where $g^{(0)}_{\mu\nu}$ and $A^{(0)}_\mu$ denote commutative quantities.
These perturbative expansions lead to the following terms in the equations of motion:
\begin{equation}\label{expcom}
\begin{split}
        -\frac{1}{2}g_{\mu\nu} \mathcal{L}(\mathcal{F},\mathcal{G}) &= -\frac{1}{2}\left(g_{\mu\nu}^{(0)} + ah_{\mu\nu}\right)\left( \mathcal{L}^{(0)} + a\mathcal{L}^{(1)}\right)\\
        &= -\frac{1}{2}g_{\mu\nu}^{(0)}\mathcal{L}^{(0)} -\frac{1}{2}a\;\left[g_{\mu\nu}^{(0)} \mathcal{L}^{(1)}+ h_{\mu\nu} \mathcal{L}^{(0)}\right]\, ,
\end{split}
\end{equation}
where
$$\mathcal{L}^{(1)} = \frac{d \mathcal{L}}{da}\Bigl\rvert_{a=0}\;.$$
\section{Perturbative solution and examples}\label{sec5}
As announced at the end of the previous section, our objective is to find the first-order perturbative solution in $a$ of the equations of motion (\ref{NC efe}) and (\ref{NC max}), where the zeroth-order solution corresponds to a commutative dyonic black hole. We will show that the NC metric corrections can be universally expressed in terms of the commutative solution, i.e. their form is independent of the underlying NLE theory. This emphasizes both the generality and simplicity of our result, as it avoids the need to solve potentially involved equations on a case-by-case basis. The solution is sought in the above-mentioned form
\begin{align}
\tensor{g}{_\mu_\nu}&=g^{(0)}_{\mu\nu}+a\tensor{h}{_\mu_\nu}\ ,\\
A_\mu&=A^{(0)}_{\mu}+aB_{\mu}
\end{align}
where $g^{(0)}_{\mu\nu}$ and $A^{(0)}_{\mu}$ are the metric and gauge potential of a commutative black hole. It should be noted that the choice of the Killing twist and the symmetry-inheriting electromagnetic field suggests considering dyonic configurations. From the SW map (\ref{SW F}), it is clear that the solutions with electric or magnetic charge only ($A_t$ or $A_\varphi$ components of gauge potential) are trivial in the sense that they coincide with the commutative ones. Namely, due to the symmetry inheritance of the electromagnetic field, we have $\tensor{\theta}{^\alpha^\beta}\partial_\beta \tensor{F}{_\mu_\nu}=0$, while the only nonvanishing contributions are proportional to $\tensor{\theta}{^t^\varphi} \tensor{F}{_t_\mu} \tensor{F}{_\varphi_\nu}$ and $\tensor{\theta}{^\varphi^t} \tensor{F}{_\varphi_\mu} \tensor{F}{_t_\nu}$, therefore only dyonic or charged rotating starting (commutative) solutions can obtain corrections in our approach. Without loss of generality, the seed metric can be put in the form
\begin{align}
g^C_{\mu\nu}=\left[\begin{array}{cccc} 
-f(r) & 0 & 0 & 0\\
0 & 1/f(r) & 0 & 0\\
0 & 0 & r^2 & 0\\
0 & 0 & 0 & r^2\sin^2\theta\\
\end{array}\right]
\end{align}
To justify the choice of this ansatz, we recall that the condition $\tensor{T}{^t_t}=\tensor{T}{^r_r}$, which is fulfilled for the NLE energy-momentum tensor, is enough to ensure that $\tensor{g}{_t_t}\tensor{g}{_r_r}=-1$ \cite{Jacobson_2007}. In the gauge sector, the Seiberg-Witten map (\ref{ASW}) constraints the form of the correction, so that the full form of gauge potential reads
\begin{align}\label{eq:NCA}
A_\mu=\left[\begin{array}{c} 
A_t(r) \\ 0 \\ 0 \\ A_\varphi(\theta) 
\end{array}\right] + a\left[\begin{array}{c} 
0 \\ -\frac{1}{2}A_\varphi\partial_rA_t \\\hspace{3mm}\frac{1}{2}A_t\partial_\theta A_\varphi \\ 0 
\end{array}\right]\;.
\end{align}
Effectively, the problem then reduces to finding the corrections to the metric tensor. \\
Before solving the equations of motion in their original form, it is useful to recognize an important simplification. Combinations of the form $\tensor{\theta}{^\alpha^\beta}\tensor{F}{_\alpha_\beta}$ and $\tensor{\theta}{^\alpha^\beta}{\ast\tensor{F}{^\alpha^\beta}}$ either vanish or are of $a^2$ order, therefore irrelevant for the lowest order corrections. This fact also reduces the number of relevant terms in the expansions of the commutative terms (\ref{expcom}). Taking these arguments into account, both Einstein's and Maxwell's equations assume more elegant forms,
\begin{align}\label{eq:NCEinstein}
\tensor{G}{_\mu_\nu}&=2(\tensor{g}{_\mu_\nu}\LL-4\LL_\mathcal{F}\tensor{F}{_\mu_\sigma}\tensor{F}{_\nu^\sigma}-\LL_\mathcal{G}\mathcal{G}\tensor{g}{_\mu_\nu}
-4\LL_\mathcal{F}\tensor{\theta}{^\alpha^\beta}(\tensor{F}{_\alpha_\mu}\tensor{F}{_\nu_\sigma}+\tensor{F}{_\mu_\sigma}\tensor{F}{_\alpha_\nu})\tensor{F}{_\beta^\sigma})
\end{align}
and
\begin{align}\label{eq:redMax}
&\partial_\mu(\sqrt{-g}(4\LL_\mathcal{F}\tensor{F}{^\mu^\nu}+4\LL_\mathcal{G}{\ast\tensor{F}{^\mu^\nu}}+4\LL_{\mathcal{F}}\tensor{\theta}{^\alpha^\beta}\tensor{g}{^\rho^\mu}\tensor{g}{^\sigma^\nu}\tensor{F}{_\alpha_\rho}\tensor{F}{_\beta_\sigma}+4\LL_{\mathcal{F}}\tensor{\theta}{^\mu^\beta}\tensor{F}{^\nu^\sigma}\tensor{F}{_\beta_\sigma}-4\LL_{\mathcal{F}}\tensor{\theta}{^\nu^\beta}\tensor{F}{^\mu^\sigma}\tensor{F}{_\beta_\sigma}+\\&+4\LL_{\mathcal{G}}\tensor{\theta}{^\mu^\beta}{\ast\tensor{F}{^\nu^\sigma}}\tensor{F}{_\beta_\sigma}-4\LL_{\mathcal{G}}\tensor{\theta}{^\nu^\beta}{\ast\tensor{F}{^\mu^\sigma}}\tensor{F}{_\beta_\sigma}+2\LL_{\mathcal{G}}\tensor{\theta}{^\alpha^\beta}\tensor{\epsilon}{^\rho^\sigma^\mu^\nu}\tensor{F}{_\alpha_\rho}\tensor{F}{_\beta_\sigma}))=0\ ,
\end{align}
respectively. Careful inspection of the equation (\ref{eq:NCEinstein}) suggests that the nontrivial novel components of metric tensor are $\tensor{h}{_t_\theta}$ and $\tensor{h}{_r_\varphi}$, which is an assumption that will be justified by further analysis. We will also show that these corrections cannot be removed by any choice of the integration constants. To solve the equations (\ref{eq:NCEinstein}) and (\ref{eq:redMax}), one can adopt a systematic approach. The $tr$ component of Einstein's equation 
\begin{align}
\cot\theta\tensor{h}{_t_\theta}f(r)'+f(r)'\partial_\theta\tensor{h}{_t_\theta}-f(r)(\cot\theta\partial_r\tensor{h}{_t_\theta}+\partial_r\partial_\theta\tensor{h}{_t_\theta})=0
\end{align}
suggests the following ansatz, $\tensor{h}{_t_\theta}=f(r)h_1(\theta)$,  while the $\theta\varphi$ component 
\begin{align}
\partial_r(-2\cot\theta f(r)\tensor{h}{_r_\varphi}+f(r)\partial_\theta\tensor{h}{_r_\varphi})=0
\end{align}
is solved by $\tensor{h}{_r_\varphi}=h_2(r)\sin^2\theta$.
Using the commutative equations of motion, the $t\theta$ and $r\varphi$ components of Einstein's equation reduce to a single equation
\begin{align}
A_\varphi'h_2(r)+2r^2A_t'A_\varphi'+r^2h_1(\theta)A_t'=0
\end{align}
whose solution may be sought in the following form
\begin{align}
h_1(\theta)=C_1A_\varphi'\ \text{and}\ h_2(r)=C_2 r^2A_t'\ ,
\end{align}
where the constants are related by $C_1=-2-C_2$. This construction automatically satisfies Maxwell's equation (\ref{eq:redMax}) up to the $a^1$ order. From the commutative Maxwell's equation it follows that $A_\varphi=-P\cos\theta$. The full solution is then given by
\begin{align}\label{eq:NCmetric}
g_{\mu\nu}=\left[\begin{array}{cccc} 
-f(r) & 0 & -a(C_2+2)f(r)P\sin\theta & 0\\
0 & 1/f(r) & 0 & aC_2r^2\sin^2\theta A_t'\\
-a(C_2+2)f(r)P\sin\theta & 0 & r^2 & 0\\
0 & aC_2r^2\sin^2\theta A_t' & 0 & r^2\sin^2\theta\\
\end{array}\right]
\end{align}
The constant $C_2$ remains undetermined as there is no obvious boundary condition that could be imposed to fix its value. However, we can consider several options; if we choose $C_2=0$, we can eliminate the ``electric" contribution to the metric correction, while for $C_2=-2$, the ``magnetic" one is absent. In any case, there is no value of $C_2$ such that both corrections are absent. 

Before proceeding with specific examples of NLE theories, several general remarks are in order. From the form of the metric (\ref{eq:NCmetric}) and the modified energy-momentum tensor (\ref{eq:NCEM}), it is obvious that the introduced NC effects generally break symmetries (conformal or SO(2) duality invariance) that may have been present in the original theory. Neither conserved charges (\ref{eq:Komar}) nor the electric field defined in (\ref{eq:fields}) are altered by the NC corrections. Notice, however, that the SW map introduces a novel nontrivial $\tensor{F}{_r_\theta}$ term in the electromagnetic tensor, which contributes to the magnetic field. To better understand the implications of the calculated corrections, in the following subsection we explicitly evaluate them within relevant NLE theories that admit known dyonic solutions. All of the considered theories are analytic at $\mathcal{F},\mathcal{G} = 0$ and as such, due to arguments in Section \ref{Subsec 3a}, (NC) gauge transformations of their corresponding Killing solutions will remain solutions, even though the theories themselves are not gauge invariant in general.

\subsection{Maxwell theory}
As an introductory example, similarly to \cite{Pozar:2025yoj}, we consider Maxwell's electrodynamics, characterized by conformal invariance and electromagnetic SO(2) duality symmetry, with Lagrangian
\begin{align}
\mathcal{L}=-\dfrac{1}{4}\mathcal{F}\ .
\end{align}
The solution describing a dyonic Reissner-Nordstr{\"{o}}m black hole is given by 
\begin{equation}
\left.\begin{aligned}
f(r)=1-\dfrac{2M}{r}+\dfrac{Q^2+P^2}{r^2}\ ,\\
A_t=-\dfrac{Q}{r},\ \ A_\varphi=-P\cos\theta \, .
\end{aligned}\right.
\end{equation}
Simply by inserting it into the metric (\ref{eq:NCmetric}) and SW map (\ref{eq:NCA}), we get the NC corrections,
\begin{equation}
\left.\begin{aligned}
\tensor{h}{_t_\theta}&=-aq(C_2+2)P\sin\theta\Big(1-\dfrac{2M}{r}+\dfrac{Q^2+P^2}{r^2}\Big),\ \ \tensor{h}{_r_\varphi}=aqC_2Q\sin^2\theta\ ,\\
A_r&=\dfrac{aqQP}{2r^2}\cos\theta,\ \ A_\theta=-\dfrac{aqQP}{2r}\sin\theta\ .
\end{aligned}\right.
\end{equation}
The NC corrected theory is no longer SO(2) duality invariant, as is noticeable from the form of the metric. At $a^1$ order, the theory is neither conformally invariant. 
\subsection{Born-Infeld theory}
Born-Infeld theory \cite{Born34, BI34} was initially proposed as a mechanism for regularizing the divergent field and self-energy of a point charge. This is achieved by imposing an upper limit on the strength of the electric field, represented by the parameter $b$ in the Lagrangian,
\begin{align}
\mathcal{L}=b^2\Bigg(1-\sqrt{1+\dfrac{\mathcal{F}}{2b^2}-\dfrac{\mathcal{G}^2}{16b^4}}\,\Bigg)\ .
\end{align}
It also emerges as a part of the effective action in the low energy limit of string theory \cite{BIstring}, where the parameter $b$ admits interpretation in terms of the string tension. To highlight a parallel with Maxwell's electrodynamics, it should be noted that it obeys the MWF limit and SO(2) electromagnetic duality invariance. The solution corresponding to a dyonic Born-Infeld black hole is 
\begin{equation}
\left.\begin{aligned}
f(r)&=1-\dfrac{2M}{r}+\dfrac{2b^2r^2}{3}-\dfrac{2b^2}{3}\sqrt{r^4+\dfrac{\tilde{Q}^2}{b^2}}+\dfrac{4\tilde{Q}^2}{3r^2}{_2F_1}\Big[\dfrac{1}{4},\dfrac{1}{2};\dfrac{5}{4};-\dfrac{\tilde{Q}^2}{b^2r^4}\Big]\ ,\\
A_t&=-\dfrac{Q}{r}{_2F_1}\Big[\dfrac{1}{4},\dfrac{1}{2};\dfrac{5}{4};-\dfrac{\tilde{Q}^2}{b^2r^4}\Big],\ \ A_\varphi=-P\cos\theta\ ,
\end{aligned}\right.
\end{equation}
where $_2F_1[a,b,c;z]$ is the hypergeometric function and $\widetilde{Q}^2=Q^2+P^2$. This example illustrates the advantage of a general approach adopted in the paper. Namely, due to the involved form of the commutative solution, solving the NC Einstein-Born-Infeld equations of motion would be computationally more challenging. Using the universal relations (\ref{eq:NCmetric}) and (\ref{eq:NCA}), we can evade this problem and evaluate the corrections, yielding
\begin{equation}
\left.\begin{aligned}
&\tensor{h}{_t_\theta}=-aq(C_2+2)P\sin\theta\Big(1-\dfrac{2M}{r}+\dfrac{2b^2r^2}{3}-\dfrac{2b^2}{3}\sqrt{r^4+\dfrac{\tilde{Q}^2}{b^2}}+\dfrac{4\tilde{Q}^2}{3r^2}{_2F_1}\Big[\dfrac{1}{4},\dfrac{1}{2};\dfrac{5}{4};-\dfrac{\tilde{Q}^2}{b^2r^4}\Big]\Big)\ ,\\
&\tensor{h}{_r_\varphi}=aqC_2r^2\sin^2\theta\dfrac{Q}{\sqrt{r^4+\tilde{Q}^2/b^2}}\ ,\\
&\tensor{A}{_r}=\dfrac{1}{2}aqP\cos\theta\dfrac{Q}{\sqrt{r^4+\tilde{Q}^2/b^2}}\ ,\ 
\tensor{A}{_\theta}=-\dfrac{1}{2}aqP\sin\theta\dfrac{Q}{r}{_2F_1}\Big[\dfrac{1}{4},\dfrac{1}{2};\dfrac{5}{4};-\dfrac{\tilde{Q}^2}{b^2r^4}\Big]\ .
\end{aligned}\right.
\end{equation}
Similarly as in the former example of Maxwell theory, the NC corrected Born-Infeld electrodynamics is no longer SO(2) duality invariant. The NC generalizations of the Dirac-Born-Infeld theory have been previously studied in \cite{Banerjee:2004rs, Aschieri:2001gf}.
\subsection{Euler-Heisenberg theory}
In their seminal paper, Euler and Heisenberg \cite{HE36} presented a one-loop QED correction to Maxwell's theory. This nonlinear interaction describes vacuum polarisation effects and may account for classically forbidden processes, such as light-by-light scattering. Its low-energy limit became known as the effective Euler-Hesienberg theory,  
\begin{align}
\mathcal{L}=-\dfrac{1}{4}\mathcal{F}+\dfrac{\alpha^2}{360m_e^4}(4\mathcal{F}^2+7\mathcal{G}^2)+\mathcal{O}(\alpha^3)\, ,
\end{align}
where $\alpha$ is the fine-structure constant and $m_e$ is the electron mass. The dyonic Euler-Heisenberg black hole, where the solution is valid up to $\alpha^2$ order, is given by
\begin{equation}
\left.\begin{aligned}
f(r)=1-\dfrac{2M}{r}+\dfrac{Q^2}{r^2}+\dfrac{P^2}{r^2}-\dfrac{4\alpha^2(P^4+5P^2Q^2+Q^4)}{225m_e^4r^6}\\
A_t=-\dfrac{Q}{r}+\dfrac{4\alpha^2Q}{45m_e^4r^5}\Big(\dfrac{2}{5}Q^2+P^2\Big),\ \ A_\varphi=-P\cos\theta
\end{aligned}\right.
\end{equation}
One can notice that the second term in the electric field falls off as $r^{-6}$ as $r\to\infty$, in accordance with the result proven in \cite{Bokulic_2024}. This is a generic feature of NLE theories obeying the MWF limit, therefore valid for Born-Infeld theory as well upon expanding the hypergeometric function.

The NC corrections are calculated straightforwardly and yield
\begin{equation}
\left.\begin{aligned}
&\tensor{h}{_t_\theta}=-aq(C_2+2)P\sin\theta\Bigg(1-\dfrac{2M}{r}+\dfrac{Q^2}{r^2}+\dfrac{P^2}{r^2}-\dfrac{4\alpha^2(P^4+5P^2Q^2+Q^4)}{225m_e^4r^6}\Bigg),\\ &\tensor{h}{_r_\varphi}=aqC_2r^2\sin^2\theta\Bigg(\dfrac{Q}{r^2}-\dfrac{4\alpha^2Q}{9m_e^4r^6}\Big(\dfrac{2}{5}Q^2+P^2\Big)\Bigg),\\
&\tensor{A}{_r}=\dfrac{1}{2}aqP\cos\theta\Bigg(\dfrac{Q}{r^2}-\dfrac{4\alpha^2Q}{9m_e^4r^6}\Big(\dfrac{2}{5}Q^2+P^2\Big)\Bigg),\ \ \tensor{A}{_\theta}=\dfrac{1}{2}aqP\sin\theta\Bigg(-\dfrac{Q}{r}+\dfrac{4\alpha^2Q}{45m_e^4r^5}\Big(\dfrac{2}{5}Q^2+P^2\Big)\Bigg)\ .
\end{aligned}\right.
\end{equation}
This example illustrates the combined effect of two types of quantum corrections - one being the NC deformation and the other manifested through the choice of the electromagnetic interaction.
\section{Discussion}\label{sec6}
In this paper we have defined noncommutative generalizations of general minimally coupled Einstein-NLE theories,
$$ S = \dfrac{1}{16\pi}\int d^4x\sqrt{-g} \left(R + 4\mathcal{L}(\mathcal{F},\mathcal{G})\right).$$
We have encountered the problem of ambiguity in the (noncommutative) ordering of index contractions in the action, which is also encountered in noncommutative Poincar\'{e} gauge theory of gravity at the level of defining metric and curvature tensors \cite{Juric:2025kjl}. Equivalently, there does not exist one preferred choice of contraction ordering, at least under the set of assumptions we have considered. The problem was circumvented by proving that in the $\partial_t\wedge \partial_\varphi$ twist, using the Palais' theorem, all contraction orderings share the same static, axially symmetric solutions. We have found a shared non-diagonal NC-corrected solution \eqref{eq:NCA},\eqref{eq:NCmetric} for any $\mathcal{L}(\mathcal{F},\mathcal{G})$ NLE theory and evaluated the obtained corrections for several prominent $\mathcal{L}$ examples. The solution reverts to the commutative solution in the commutative limit $a\rightarrow 0$.\\

Our discussion focused on the Killing scenario defined by the $\partial_t \wedge \partial_\varphi$ twist, and static, axially symmetric solutions. We have shown that gauge invariance is preserved within this particular setup. In general, however, a simple replacement of ordinary commutative products by star products, combined with the SW map for the gauge fields, does not lead to an NC gauge invariant action. In principle, one could also consider the semi-Killing case in which the metric and the gauge potential are invariant with respect to only one of the vector fields entering the twist. This could be utilised to study static noncommutative spacetimes which are not axially symmetric in $\partial_t\wedge \partial_r$ twist, or to calculate (axially symmetric) time evolution of black holes in the $\partial_\varphi \wedge \partial_r$ twist. However, with the caveat that gauge invariance of the solution is lost as soon as one moves away from the Killing twist.\\

There are several universal conclusions which hold for all NC deformed solutions \eqref{eq:NCA}, \eqref{eq:NCmetric}. Namely, the mass $M$, the absence of the angular momentum, electric charge $Q$ and magnetic charge $P$ all remain unchanged for our solutions. The same holds true for the Ricci scalar curvature $R$ as well as $R_{\mu\nu}R^{\mu\nu}$ and $R_{\mu\nu\rho\sigma} R^{\mu\nu\rho\sigma}$. Additionally, all of the metrics \eqref{eq:NCmetric} depend on an unknown constant $C_2$ appearing in the new off-diagonal terms. One can see that, in agreement with our claims that corrections exist only for spaces with nonzero electric and magnetic field, it is not possible to choose $C_2$ such that both $g_{t\theta}$ and $g_{r\varphi}$ vanish. Finally, from the form of the novel non-diagonal metric terms, it is evident that the metric \eqref{eq:NCmetric} is not asymptotically Minkowski.\\

Regarding the future research directions, it would also be interesting to see if the NC corrections to the NLE Lagrangian truncate after $a^2$ order, as they do in \cite{DimitrijevicCiric:2024ibc} for the Maxwell electrodynamics. In that case, the action would be exact, rather than just a perturbative expansion. It is worth noting that, just as is the case in \cite{Pozar:2025yoj}, the metric of the solution \eqref{eq:NCmetric} is up to a prefactor the effective metric\footnote{Which is obtained from just an algebraic system of equations, not differential equations.} in the sense of \cite{DimitrijevicCiric:2024ibc}. It should be investigated if this pattern continues in higher orders of $a$ and for theories with different matter contents. This finding, if it turns out to be true, would be important because the $a^2$ expanded action \eqref{NC NLE} leads to nonlinear equations of motion, and using the effective metric as an ansatz would substantially simplify the process of solving them. Finally, our method should also apply to actions with modified gravitational sector coupled to NLE. There, the Killing twist will again render the LHS of the (modified) Einstein equation unchanged, with the NC corrections arising from the matter sector. In the commutative case, previous analyses have primarily focused on electrically charged black holes with NLE sources, for example, within Einstein-Gauss-Bonnet gravity \cite{PhysRevD.83.064017, PhysRevD.83.024011} or $f(R)$ theory \cite{PhysRevD.78.124007}. However, such solutions will not acquire the NC corrections under the Killing twist deformation considered in this work. Therefore, the starting point would be the construction of dyonic (or, more ambitiously, Kerr-Newman configurations) in these theories.

\acknowledgments
Both authors thank Ivica Smoli\'c for suggestions and reading the first draft prior to publication. The authors also thank Tajron Juri\'{c} for his valuable comments.
This research was supported by the Croatian Science Foundation Project No. IP-2025-02-8625, "Quantum aspects of gravity". A. B. is further supported by CIDMA under the Portuguese Foundation for
Science and Technology (FCT, https://ror.org/00snfqn58) Multi-Annual
Financing Program for R\&D Units,grants UID/4106/2025 and UID/PRR/
4106/2025, as well as the projects: Horizon Europe staff exchange (SE) programme HORIZON-MSCA2021-SE-01 Grant No. NewFunFiCO-101086251;  2022.04560.PTDC (\url{https://doi.org/10.54499/2022.04560.PTDC}) and 2024.05617.CERN (\url{https://doi.org/10.54499/2024.05617.CERN}).\\

\bibliography{BibTex}

\end{document}